\documentclass[prl,twocolumn,showpacs,amsmath,amssymb,superscriptaddress,  nofootinbib]{revtex4-1}

\usepackage{graphicx}% Include figure files
\usepackage{dcolumn}% Align table columns on decimal point
\usepackage{bm}     % bold math
\usepackage{multirow}
\usepackage{booktabs}
\usepackage{amsmath}
\usepackage{braket}
\usepackage{appendix}
\usepackage{float}
\usepackage{xcolor}
\usepackage{siunitx}
\usepackage{amssymb}
\usepackage{array}
\usepackage{hyperref}

\begin{document}
	\title {Reentrant Bloch ferromagnetism}
	\author{DinhDuy Vu}
	\affiliation{Condensed Matter Theory Center and Joint Quantum Institute, Department of Physics, University of Maryland, College Park, Maryland 20742, USA}
	
	\author{S. Das Sarma}
	\affiliation{Condensed Matter Theory Center and Joint Quantum Institute, Department of Physics, University of Maryland, College Park, Maryland 20742, USA}
	
	\begin{abstract}
		An interacting electron liquid in two (2D) and three (3D) dimensions may undergo a paramagnetic-to-ferromagnetic quantum spin polarization transition at zero applied magnetic field, driven entirely by exchange interactions, as the system density ($n$) is decreased. This is known as Bloch ferromagnetism. We show theoretically that the application of an external magnetic field ($B$), which directly spin polarizes the system through Zeeman spin splitting, has an interesting effect on Bloch ferromagnetism if the applied field and carrier density are both decreased (from some initial applied high magnetic field at a high carrier density) in a power-law manner, $B\sim n^p$. For $p<p_c$, with $p_c= 1 (2/3)$ in $2(3)$D, the system remains either fully spin-polarized or undergoes a single transition from a partially spin-polarized (with two Fermi surfaces corresponding to spin up and down electrons) to a fully spin-polarized state (with a single Fermi surface of one spin) as the density and field decrease, depending on whether the starting point is partially spin-polarized or fully spin-polarized. However, for $p>p_c$, the system may undergo two transitions if starting from the fully spin-polarized state: first, a weak second order transition at high density and field from the field-induced fully polarized phase to the partially polarized phase; and then, at a lower field and density, a reentrant first order transition back to the fully spin-polarized phase again with a single Fermi surface.
	
	\end{abstract}
	\maketitle
	\textit{Introduction -} In 1927, Bloch made an interesting prediction \cite{Bloch1929}, which could be construed as the first theoretical development in quantum many body theories as well as among the first concrete quantum predictions in solid state physics. He pointed out that as the carrier density is decreased, an interacting electron gas (in a uniform positively charged background for stability and charge neutrality) undergoes a phase transition at a critical density from being a higher-density paramagnetic spin unpolarized system with equally populated Fermi surfaces of both spins to a lower-density spin polarized ferromagnetic system with only one Fermi surface of same spins. The physics behind this is the exchange energy arising from the combination of Coulomb interaction and Pauli principle, leading to the system minimizing its energy at low densities through spin polarization since this optimizes the exchange contribution by making the electrons stay far away from each other by virtue of the Pauli principle. A noninteracting electron gas obviously has a paramagnetic ground state in order to minimize the kinetic energy by populating both spin states equally to ensure a low Fermi energy. Bloch showed that the exchange-driven transition from the paramagnetic to the ferromagnetic state occurs sharply at a low critical density. In 3D metals, the Bloch ferromagnetism occurs at a far too low carrier density to be of any physical relevance, but in a 2D semiconductor system, where the carrier density can be tuned by an external gate voltage, Bloch ferromagnetism has been much studied because, in principle, the carrier density can be made low enough to access Bloch ferromagnetism \cite{Rajagopal1977,Rajagopal1978,Yarlagadda1989,Zhang2005,Attaccalite2002,Drummond2009}.
	
    We emphasize that our theorem for the existence of a magnetic field-induced reentrant ferromagnetic transition is completely general and independent of the specific Hartree-Fock formalism we use, which is done for analytical transparency. We are establishing a generic principle in this work valid beyond the details of the Hartree-Fock theory. Other more numerically accurate approaches exhibiting Bloch ferromagnetism such as RPA \cite{Rajagopal1977,Rajagopal1978,Yarlagadda1989,Zhang2005} and Quantum Monte Carlo \cite{Attaccalite2002,Drummond2009} would only modify the position of the critical point, manifesting the same universal reentrant transition in an applied field. 
	
	Motivated by a recent experiment \cite{R3}, where Bloch ferromagnetism has been reported in 2D composite fermions, we study theoretically the interplay between an applied magnetic field, coupled to electron spins causing the usual field-induced Zeeman spin splitting, and carrier density-induced exchange coupling leading to Bloch type spin-polarization in 2D electrons (We also provide results for the corresponding 3D case for completeness.). We consider a situation where the applied magnetic field $B$ changes, to be consistent wit the experiment in Ref.~\cite{R3}, along with a change in density $n$ according to a power-law specified manner: $B \sim n^p$, where $p$ is a power law exponent. In the composite fermion experiment $p=1$ \cite{R3}, but we consider arbitrary $p$, asking how the ground state spin polarization of the interacting electron liquid changes as $B, n$ both are varied together. In the experiment \cite{R3}, $B\sim n$ was necessary simply to keep the 2D electron system maintain a fixed Landau level filling of half so that the composite fermions themselves, carrying two effective units of average flux quanta, can be thought of as being at zero magnetic field according to the well-known Jain prescription \cite{Jain1989,*Park1998,*Park1999,Halperin1993}. In this Letter, we consider real electrons in a Zeeman field with varying field/density according to $B \sim n^p$ to study how the spin-polarizatin of the system evolves under the combined effect of varying density and field in a concomitant manner. In particular, one goal of our work is to study Bloch ferromagnetism in the presence of external Zeeman field producing direct spin-polarization.

   The experiment we have in mind is a 2D (or 3D, we provide results for both cases) electron gas in a parallel magnetic field causing some spin-polarization initially, and then varying both the magnetic field and the carrier density together as $B \sim n^p$ to observe their combined effects on the net spin-polarization (i.e. including both the magnetic field and the interacting exchange energy). This is qualitatively the same as the recent composite fermion experiment with the only qualitative difference being that we are considering fundamental fermions (i.e. usual electrons with no flux tubes attached) in contrast to the composite fermions. Our work on regular electrons, compared to the experiment on composite fermions, sheds light on an important old question of how real and physical the composite fermions are in addition to discerning the nature of Bloch ferromagnetism in an applied magnetic field.
		
  The Zeeman field by itself obviously produces full spin polarization even for noninteracting electrons when the Zeeman splitting between up and down electrons surpasses the Fermi energy so that the system has only one Fermi surface of the same spin at high field (and two equal Fermi surfaces of opposite spins at zero field). Reducing electron density by itself may also lead to a single spin-polarized Fermi surface in zero field in an interacting system by virtue of Bloch ferromagnetism. The two together should therefore hasten the onset of full spin polarization, but the experimental question we address is subtle. As the density decreases increasing exchange effects, the field strength decreases proportionally, leading to the possibility that an initially field-polarized state may become unpolarized or remain spin-polarized depending on the details. What are the constraints determining this behavior? Experimentally, where $B \sim n$, the composite fermions were reported to be initially completely spin-polarized (i.e. one Fermi surface), becoming partially spin-polarized first (i.e. two Fermi surfaces), and then undergoing a second transition to complete spin polarization again (i.e. back to just one Fermi surface) in a second transition at a still lower field. Do regular electrons follow the same behavior? Are there constraints on the exponent $p$ (=1 in the composite fermion experiment) leading to such a reentrant ferromagnetic transition (1-2-1 Fermi surfaces) or is this always guaranteed for any $p$?

  We find, as described below, that there is a critical $p=p_c$ determining whether there are two transitions ($p>p_c$) or no transition (for a starting spin-polarized state) or one transition (for an initially spin partially polarized state). We theoretically establish that $p_c=1 (2/3)$ for 2(3)D systems, and thus reentrant Bloch ferromagnetism arises only if the applied magnetic field is changing (at least infinitesimally) faster than linear in 2D systems. It is allowed for the system not to show any transition at all (for any $p$), remaining fully trivially spin-polarized with decreasing field/density depending on the starting point, e.g., if the starting point is at a very high applied field the system remains spin-polarized throughout never becoming partially spin-polarized and there is no transition from one to two Fermi surfaces for any $p$ in such a situation, and thus $p_c$ separates behaviors involving two or no transitions (for $p>p_c$) from one or no transition (for $p<p_c$).  Interestingly, the experiment of Ref.~\cite{R3} involves the marginal value of $p=p_c$ since $B\sim n$.
		
  \textit{Theory -} Since our interest is a matter of principle and not quantitative estimates of critical density or magnetic field, we use the Hartree-Fock theory and express all quantities in dimensionless units. It is well-known that the Hartree-Fock theory, which is fully analytical (thus enabling us to obtain unambiguous answers to the questions addressed in this work), gives the same qualitative results as more accurate (and completely numerical) approximations (e.g. quantum Monte Carlo) except for the critical density for the ferromagnetic transition density which is over-estimated in the Hartree-Fock theory. 
  
  Our starting Hamiltonian is an electron gas of total density $n= n_\uparrow + n_\downarrow$, where $\uparrow/\downarrow$ refer to the two spin states with the standard quadratic noninteracting kinetic energy dispersion and the long-range Coulomb interaction $e^2/r$ among the electrons. We describe the energy in terms of the Hartree energy $E_h=\hbar^2/(ma_B^2)$ with $a_B$ being the Bohr radius. We also add a Zeeman splitting term $b=\mu g B/E_h$ introduced by the applied magnetic field $B$. The total ground state energy per electron can be obtained in 2D and 3D as closed expressions (the three terms in the energy are respectively the kinetic energy, exchange energy, and Zeeman energy) in terms of electron density and magnetization $\xi= (n_\uparrow-  n_\downarrow)/n$ as:
  	\begin{align}
  	E_{2D}&=\frac{1+\xi^2}{2r_s^2} - \frac{2\sqrt{2}\left[(1+\xi)^{3/2}+(1-\xi)^{3/2}\right]}{3\pi r_s} - b \xi \label{eq1}\\
    \begin{split}
    E_{3D}&= \frac{3\cdot(9\pi)^{2/3}}{5\cdot 2^{10/3}}\frac{(1+\xi)^{5/3}+(1-\xi)^{5/3}}{r_s^2}\\
    &\quad- \frac{3^{5/3}}{2^{11/3}\pi^{2/3}}\frac{(1+\xi)^{4/3}+(1-\xi)^{4/3}}{r_s} - b \xi
    \end{split}\label{eq2}
    \end{align}
  Here $r_s$ is the usual dimensionless inter-electron separation (or Wigner-Seitz radius) characterizing the interaction strength: 
  $r_s= 2(k_Fa_B)^{-1}$	(2D), or $(9\pi/2)^{1/3}(k_Fa_B)^{-1}$  (3D) with $k_F \sim n^{1/2} (n^{1/3})$ in 2(3)D.
  
  Through minimizing the energy, we obtain the magnetization defining the paramagnetic and ferromagnetic phases with $\xi <1$ (two Fermi surfaces) and $\xi=1$ (one Fermi surface), respectively. Within the Hartree-Fock approximation, the phase boundary in 2D can be worked out analytically through the parametric equations (similar but much more lengthy expressions for 3D are not shown here)
  \begin{align}
  	r_s(\Xi) &= \frac{3\pi\Xi_-^2}{16-8\sqrt{2}\Xi_+^{1/2}+2\sqrt{2}\Xi_- \left(\Xi_+^{1/2}-\Xi_-^{1/2}\right)}, \label{eq3} \\
    b(\Xi) &= r_s(\Xi)^{-2}\left[ \Xi- \frac{\sqrt{2}}{\pi}\left( \Xi_+^{1/2}-\Xi_-^{1/2}\right)r_s(\Xi)\right]. \label{eq4}
  \end{align}  
  Here, $\Xi\in[0,1]$ and $\Xi_{\pm}=1\pm\Xi$. Physically, $1-\Xi$ is the magnetization jump at the transition point. The Bloch transition critical $r_{sc}$ is computed by substituting $\Xi=0$ into Eq.~\eqref{eq3} to get $r_{sc} \approx 2.0$. Similarly, at the limit $\Xi=1$, the non-interacting Zeeman transition is described by $b_c=1/r_s^2$.

\begin{figure}
  \centering
  \includegraphics[width=0.48\textwidth]{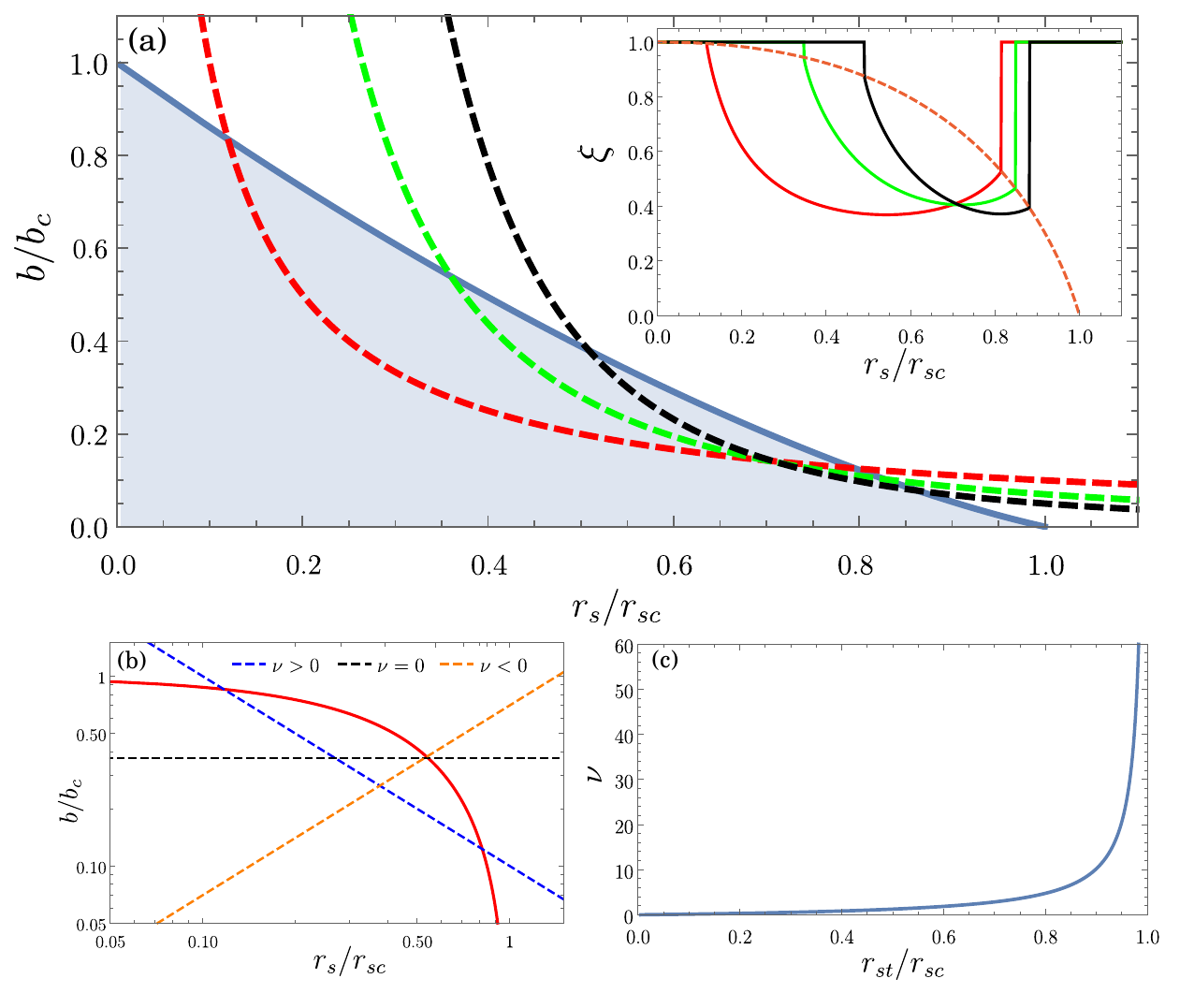}
  \caption{(a) Two-dimensional ferromagnetic (unshaded) and paramagnetic (shaded) phases in the scale of normalized $b/b_c$ and $r_s/r_{sc}$. Here, $b_c\sim 1/r_s^2$ is the critical non-interacting field and $r_{sc}$ is the critical point of field-free Bloch transition. The dashed lines exemplifies some variation processes $b\sim r_s^{-(2+\nu)}$ with $\nu=1$ (red), 2 (green), 3 (black). The inset shows the magnetization $\xi$ along each variation line corresponding color-wise to the main figure. The orange dashed line represents $\Xi$ with $1-\Xi$ being the magnetization jump at each transition point. (b) The phase boundary (solid line) expressed in the log-log scale, and three cases of power laws  (dashed lines). Only $\nu>0$ can generate reentrant ferromagnetism. (c) The tangent point $r_{st}$ between the phase boundary line and a class of power-law functions characterized by $b\sim r_s^{-(2+\nu)}$.\label{fig1}}
  \end{figure}

  In Fig.~\ref{fig1}(a), we plot the phase diagram containing the partially and fully spin-polarized phases in the $(b,r_s)$ space with $r_s$ normalized by the Bloch critical density $r_{sc}$ and Zeeman energy $b$ normalized by the non-interacting critical field $b_c$ for full polarization. In the same diagram, we plot several power-law curves $b\sim r_s^{2+\nu}$ which cut the phase boundary at two distinct points. Along these curves starting from low $r_s$ (i.e. high density), one first enters the partially polarized phase (with two Fermi surfaces) from the fully polarized one (with one Fermi surface) then re-enters back to the ferromagnetic phase at higher $r_s$ [see the inset of Fig.~\ref{fig1}(a)]. We note that except for $r_s=0$, all the transition points along the phase boundary are first-order with non-zero magnetization jumps; however, this discontinuity is weak for low $r_s$ and more significant for $r_s\to r_{sc}$, making the first transition at high field and density weakly second-order while the second one at lower field and density clearly first-order.   Thus, experimentally, the higher-field transition from one Fermi surface to two would appear to be second order, and the lower field reentrant transition back to one Fermi surface again would appear to be first-order.
  
  Even though we demonstrate the reentrance analytically within Hartree-Fock approximation, its existence can be figured out generically for any situation provided the theory predicts a low-density (high-$r_s$) Bloch ferromagnetic transition at zero applied field. The non-interacting purely field-induced transition must occur for $b\sim E_F\sim 1/r_s^2$, now if the field-free Bloch transition also exists, then the boundary line must be compact in the $(br_s^2,r_s)$ scale. Therefore, any functions $b\sim r_s^{-(2+\nu)}$ with $\nu>0$ can potentially cut the boundary line at two distinct points while for $\nu \le 0$, only one intersection at most is allowed [Fig.~\ref{fig1}(b)]. We emphasize that this necessary condition is a rigorous theorem (independent of our Hartree-Fock theory) and  is completely established from the compactness of the phase transition (or the existence of Bloch transition) and the quadratic dispersion of the kinetic energy, and thus should be valid regardless of the approximation model. The sufficient condition, on the other hand, is model-dependent. Indeed, for $\nu>0$, one can observe reentrant ferromagnetism or no transition at all depending on the coefficient in front of the power law. In Fig.~\ref{fig1}(c), we study the position of the tangent point $r_{st}$ with respect to $~\nu$ within the Hartree-Fock approximation. The exact position of the tangent point may vary with the theoretical methods, but some qualitative conclusion can be made. For $\nu\to 0$, $r_{st}\to 0$ while for $\nu\to \infty$, $r_{st}\to r_{sc}$; as the positions of the two transitions are located on the two sides of $r_{st}$, in both limits of low and high $\nu$, the intermediate paramagnetic phase may be very narrow, making it difficult to observe experimentally. We note that for $\nu\le0$, it is also possible for there to be no transition if one starts from a very high initial field so that the system remains spin-polarized always.  Thus, the two possibilities depending on the value of $\nu$ are two or zero transitions and one or zero transition.  The interesting reentrance is only allowed for $\nu>0$, but may not actually happen if the starting point is deep into the Zeeman-polarized phase. 
  
\begin{figure}
	\centering
	\includegraphics[width=0.48\textwidth]{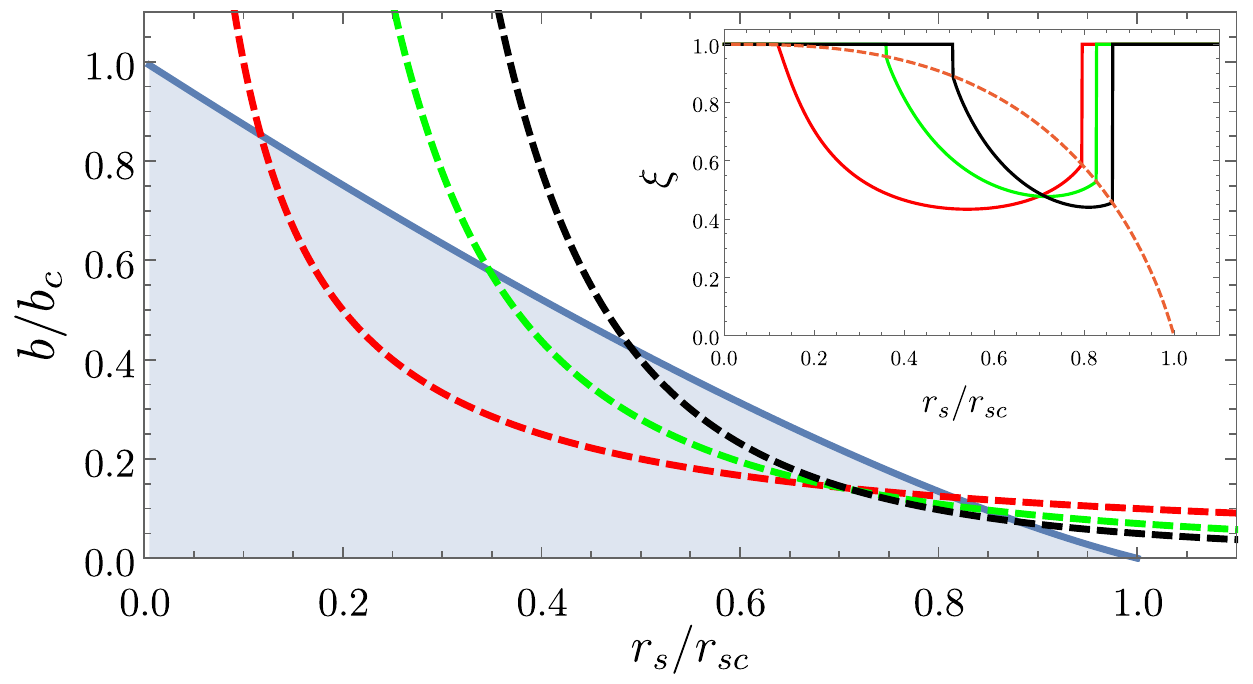}
	\caption{Three-dimensional ferromagnetic and paramagnetic phases. The notations and variation lines are similar to Fig.~\ref{fig1}(a). The inset shows the magnetization transition along each line. The results are relatively similar to the 2D case.\label{fig2}}
 \end{figure}

  The previous calculations can also be straightforwardly performed for the 3D case starting from Eq.~\eqref{eq2}. We can obtain the parametric equations for the 3D phase transition but do not show them in this Letter due to their lengthy forms. Within the Hartree-Fock approximation, the 3D field-free Bloch and non-interacting Zeeman transitions occur at $r_{sc}\approx5.5$ and $b_c \approx 1.5r_s^{-2}$. In Fig.~\ref{fig2}, we plot the 3D phase diagram in the properly normalized $b$ and $r_s$ and demonstrate power-law curves similar to the 2D case. The results show no significant discrepancy, confirming the general validity of the reentrant ferromagnetism in the presence of Bloch transition. The difference between dimensionalities only emerges when we translate the necessary condition in dimensionless variables $b\sim r_s^{-(2+\nu)}$ with $\nu>0$ to the one in physical variables $B \sim n^p$. Specifically, for 2D, $r_s\sim n^{-1/2}$ resulting in $p>1$; while for 3D, $r_s\sim n^{-1/3}$ leading to $p>2/3$.
  
  \textit{Discussion and Conclusion -} Our key result is that the existence or not of a reentrant Bloch ferromagetism depends crucially on the density variation of the applied magnetic field being larger than linear in 2D (larger than 2/3 power in 3D), and if such a reentrant double transition (one to two to one Fermi surface) takes place, then the higher (lower) density and field transition is always weakly second (first) order. If the density variation with the magnetic field is chosen to be weaker than the critical value (i.e. $p<p_c=1$, $2/3$ for 2D, 3D respectively), then an initial fully spin-polarized state remains spin-polarized (just one Fermi surface) without any transition and an initially partially spin polarized state (two Fermi surfaces) undergoes a single transition to a fully spin-polarized state (one Fermi surface). Our predictions should be verifiable in 2D electron systems such as 2D GaAs structures by varying density and applied field. The existence of a critical exponent $p_c$ and its values in 2D/3D are not dependent on our use of the analytical Hartree-Fock theory which exhibits Bloch ferromagnetism would manifest this behavior, and our qualitative results are of generic validity. The predicted $p_c$ would not change even if the interaction energy is modified as long as the kinetic energy dispersion is quadratic (and Bloch transition exists for zero applied field).
  
  Finally, in connecting our work to the recent 2D composite fermion experiment \cite{R3} which serves as our motivation, we note that nominally this experiment has $p=1$ which is the marginal case in the theory, thus strictly just (barely) ruling out a reentrant Bloch transition. This finding of ours is in fact consistent with the theoretical numerical Monte Carlo results presented in this paper \cite{R3} where the energetics also almost rule out the reentrant transition reported in the experiment. In agreement with our theory, however, the observed transition is weakly second-order (first order) at higher (lower) field and density for the composite fermions. Why does the experiment find an apparent reentrant transition even when $B$ is being tuned linearly in $n$? There are two possibilities we can think of: (1) although the variation of $B$ in $n$ is linear in the experiment, the actual variation of the Zeeman energy (i.e. $b$ in our theory) is still slightly super-linear because $g$ itself has a weak density dependence; (2) the energy dispersion of composite fermions is sufficiently different from quadratic rendering our theory inapplicable. We mention, however, that any simple nonparabolicity with cubic or quartic momentum dependent correction to the parabolic energy dispersion would increase the 2D critical exponent $p_c$ to 3/2 or 2 respectively, whereas a linear Dirac-like band dispersion eliminates the Bloch transition. We do note that the experimental observation of the intermediate paramagnetic phase is rather fragile in the experiment lasting over a very small density regime, which is consistent with our finding of the linear field-density relationship being marginal (i.e. $p_c=1$), and it is entirely possible that the experimental exponent $p$ is very slightly above 1, making our predicted reentrant ferromagnetism just feasible in the experiment. It is remarkable that our Hartree-Fock theory for the usual electrons has such a remarkable agreement with the composite fermions carrying flux tubes in a marginal Fermi liquid, justifying the original intuition \cite{Jain1989,*Park1998,*Park1999} that composite fermions behave for all practical purposes just like regular fermions by absorbing appropriate numbers of flux quanta.
   	
  We conclude by stating that we have established a theorem that a reentrant Bloch ferromagnetic transition may occur in interacting electron liquids in the presence of an applied Zeeman field varying as $n^p$ power of carrier density only if the exponent $p>p_c$, where the marginal exponent $p_c=1~(2/3)$ in 2D (3D), with the reentrant transition at lower density and field always being (weakly) first-order (and any higher field transition being weakly second order).  This universal critical exponent is derived from $b_c\sim r_s^{-2}$ in the limit $r_s\to 0$, where the Hartree-Fock treatment is asymptotically exact. Thus, our calculated  exponent value and the re-entrance phenomenon itself remain universally valid beyond our approximation scheme, but the detailed $r_s$ values defining the re-entrance in our theory are inaccurate since the Hartree-Fock approximation overestimates the importance of Bloch ferromagnetism.  We emphasize that although we cannot predict whether the interaction-driven ferromagnetic transition occurs or not in an interacting electron liquid (because we use the Hartree-Fock theory), we can assert with confidence that if Bloch ferromgmetism does occur in an experiment, our predicted re-entrance is guaranteed to happen defined by our calculated exponent.  In fact, most numerically sophisticated many body calculations \cite{Attaccalite2002,Drummond2009} beyond the Hartree-Fock theory predict the existence of Bloch ferrmagnetism, albeit at much higher $r_s$ than our Hartree-Fock predictions, and our theorem about the existence of a re-entrant transition would be valid in all these situations (except at much higher $r_s$ values).  We note in this context that the Hartree-Fock approximation is a conserving leading-order many-body approximation, in the Baym-Kadanoff sense obeying the Ward identity, as it involves only the leading order exchange self-energy diagram, without any vertex correction, using the Coulomb interaction. This makes our result of great physical interest, particularly since Bloch ferromagnetism has already been reported experimentally in composite fermions, thus making our theory physically relevant not only for ordinary electrons, but also for composite fermions. We have also studied the 1D electron system, which in the presence of interactions, becomes a Luttinger liquid with no Fermi surface, and consequently no Bloch ferromagnetism in the 1D system because of its singular Luttinger liquid nature.
   	
  \textit{Acknowledgement -} We thank Jainendra Jain and Mansour Shayegan for discussions. This work is supported by the Laboratory for Physical Sciences. 
  
	\bibliographystyle{apsrev4-1}
    \bibliography{reference}

\begin{thebibliography}{12}%
\makeatletter
\providecommand \@ifxundefined [1]{%
 \@ifx{#1\undefined}
}%
\providecommand \@ifnum [1]{%
 \ifnum #1\expandafter \@firstoftwo
 \else \expandafter \@secondoftwo
 \fi
}%
\providecommand \@ifx [1]{%
 \ifx #1\expandafter \@firstoftwo
 \else \expandafter \@secondoftwo
 \fi
}%
\providecommand \natexlab [1]{#1}%
\providecommand \enquote  [1]{``#1''}%
\providecommand \bibnamefont  [1]{#1}%
\providecommand \bibfnamefont [1]{#1}%
\providecommand \citenamefont [1]{#1}%
\providecommand \href@noop [0]{\@secondoftwo}%
\providecommand \href [0]{\begingroup \@sanitize@url \@href}%
\providecommand \@href[1]{\@@startlink{#1}\@@href}%
\providecommand \@@href[1]{\endgroup#1\@@endlink}%
\providecommand \@sanitize@url [0]{\catcode `\\12\catcode `\$12\catcode
  `\&12\catcode `\#12\catcode `\^12\catcode `\_12\catcode `\%12\relax}%
\providecommand \@@startlink[1]{}%
\providecommand \@@endlink[0]{}%
\providecommand \url  [0]{\begingroup\@sanitize@url \@url }%
\providecommand \@url [1]{\endgroup\@href {#1}{\urlprefix }}%
\providecommand \urlprefix  [0]{URL }%
\providecommand \Eprint [0]{\href }%
\providecommand \doibase [0]{http://dx.doi.org/}%
\providecommand \selectlanguage [0]{\@gobble}%
\providecommand \bibinfo  [0]{\@secondoftwo}%
\providecommand \bibfield  [0]{\@secondoftwo}%
\providecommand \translation [1]{[#1]}%
\providecommand \BibitemOpen [0]{}%
\providecommand \bibitemStop [0]{}%
\providecommand \bibitemNoStop [0]{.\EOS\space}%
\providecommand \EOS [0]{\spacefactor3000\relax}%
\providecommand \BibitemShut  [1]{\csname bibitem#1\endcsname}%
\let\auto@bib@innerbib\@empty
%</preamble>
\bibitem [{\citenamefont {Bloch}(1929)}]{Bloch1929}%
  \BibitemOpen
  \bibfield  {author} {\bibinfo {author} {\bibfnamefont {F.}~\bibnamefont
  {Bloch}},\ }\href {\doibase 10.1007/BF01340281} {\bibfield  {journal}
  {\bibinfo  {journal} {Z. Phys.}\ }\textbf {\bibinfo {volume} {57}},\ \bibinfo
  {pages} {545} (\bibinfo {year} {1929})}\BibitemShut {NoStop}%
\bibitem [{\citenamefont {Rajagopal}\ and\ \citenamefont
  {Kimball}(1977)}]{Rajagopal1977}%
  \BibitemOpen
  \bibfield  {author} {\bibinfo {author} {\bibfnamefont {A.~K.}\ \bibnamefont
  {Rajagopal}}\ and\ \bibinfo {author} {\bibfnamefont {J.~C.}\ \bibnamefont
  {Kimball}},\ }\href {\doibase 10.1103/PhysRevB.15.2819} {\bibfield  {journal}
  {\bibinfo  {journal} {Phys. Rev. B}\ }\textbf {\bibinfo {volume} {15}},\
  \bibinfo {pages} {2819} (\bibinfo {year} {1977})}\BibitemShut {NoStop}%
\bibitem [{\citenamefont {Rajagopal}\ \emph {et~al.}(1978)\citenamefont
  {Rajagopal}, \citenamefont {Singhal}, \citenamefont {Banerjee},\ and\
  \citenamefont {Kimball}}]{Rajagopal1978}%
  \BibitemOpen
  \bibfield  {author} {\bibinfo {author} {\bibfnamefont {A.~K.}\ \bibnamefont
  {Rajagopal}}, \bibinfo {author} {\bibfnamefont {S.~P.}\ \bibnamefont
  {Singhal}}, \bibinfo {author} {\bibfnamefont {M.}~\bibnamefont {Banerjee}}, \
  and\ \bibinfo {author} {\bibfnamefont {J.~C.}\ \bibnamefont {Kimball}},\
  }\href {\doibase 10.1103/PhysRevB.17.2262} {\bibfield  {journal} {\bibinfo
  {journal} {Phys. Rev. B}\ }\textbf {\bibinfo {volume} {17}},\ \bibinfo
  {pages} {2262} (\bibinfo {year} {1978})}\BibitemShut {NoStop}%
\bibitem [{\citenamefont {Yarlagadda}\ and\ \citenamefont
  {Giuliani}(1989)}]{Yarlagadda1989}%
  \BibitemOpen
  \bibfield  {author} {\bibinfo {author} {\bibfnamefont {S.}~\bibnamefont
  {Yarlagadda}}\ and\ \bibinfo {author} {\bibfnamefont {G.~F.}\ \bibnamefont
  {Giuliani}},\ }\href {\doibase 10.1103/PhysRevB.40.5432} {\bibfield
  {journal} {\bibinfo  {journal} {Phys. Rev. B}\ }\textbf {\bibinfo {volume}
  {40}},\ \bibinfo {pages} {5432} (\bibinfo {year} {1989})}\BibitemShut
  {NoStop}%
\bibitem [{\citenamefont {Attaccalite}\ \emph {et~al.}(2002)\citenamefont
  {Attaccalite}, \citenamefont {Moroni}, \citenamefont {Gori-Giorgi},\ and\
  \citenamefont {Bachelet}}]{Attaccalite2002}%
  \BibitemOpen
  \bibfield  {author} {\bibinfo {author} {\bibfnamefont {C.}~\bibnamefont
  {Attaccalite}}, \bibinfo {author} {\bibfnamefont {S.}~\bibnamefont {Moroni}},
  \bibinfo {author} {\bibfnamefont {P.}~\bibnamefont {Gori-Giorgi}}, \ and\
  \bibinfo {author} {\bibfnamefont {G.~B.}\ \bibnamefont {Bachelet}},\ }\href
  {\doibase 10.1103/PhysRevLett.88.256601} {\bibfield  {journal} {\bibinfo
  {journal} {Phys. Rev. Lett.}\ }\textbf {\bibinfo {volume} {88}},\ \bibinfo
  {pages} {256601} (\bibinfo {year} {2002})}\BibitemShut {NoStop}%
\bibitem [{\citenamefont {Zhang}\ and\ \citenamefont {{Das
  Sarma}}(2005)}]{Zhang2005}%
  \BibitemOpen
  \bibfield  {author} {\bibinfo {author} {\bibfnamefont {Y.}~\bibnamefont
  {Zhang}}\ and\ \bibinfo {author} {\bibfnamefont {S.}~\bibnamefont {{Das
  Sarma}}},\ }\href {\doibase 10.1103/PhysRevB.72.115317} {\bibfield  {journal}
  {\bibinfo  {journal} {Phys. Rev. B}\ }\textbf {\bibinfo {volume} {72}},\
  \bibinfo {pages} {115317} (\bibinfo {year} {2005})}\BibitemShut {NoStop}%
\bibitem [{\citenamefont {Drummond}\ and\ \citenamefont
  {Needs}(2009)}]{Drummond2009}%
  \BibitemOpen
  \bibfield  {author} {\bibinfo {author} {\bibfnamefont {N.~D.}\ \bibnamefont
  {Drummond}}\ and\ \bibinfo {author} {\bibfnamefont {R.~J.}\ \bibnamefont
  {Needs}},\ }\href {\doibase 10.1103/PhysRevLett.102.126402} {\bibfield
  {journal} {\bibinfo  {journal} {Phys. Rev. Lett.}\ }\textbf {\bibinfo
  {volume} {102}},\ \bibinfo {pages} {126402} (\bibinfo {year}
  {2009})}\BibitemShut {NoStop}%
\bibitem [{\citenamefont {Hossain}\ \emph {et~al.}(2021)\citenamefont
  {Hossain}, \citenamefont {Zhao}, \citenamefont {Pu}, \citenamefont {Mueed},
  \citenamefont {Ma}, \citenamefont {{Villegas Rosales}}, \citenamefont
  {Chung}, \citenamefont {Pfeiffer}, \citenamefont {West}, \citenamefont
  {Baldwin}, \citenamefont {Jain},\ and\ \citenamefont {Shayegan}}]{R3}%
  \BibitemOpen
  \bibfield  {author} {\bibinfo {author} {\bibfnamefont {M.~S.}\ \bibnamefont
  {Hossain}}, \bibinfo {author} {\bibfnamefont {T.}~\bibnamefont {Zhao}},
  \bibinfo {author} {\bibfnamefont {S.}~\bibnamefont {Pu}}, \bibinfo {author}
  {\bibfnamefont {M.~A.}\ \bibnamefont {Mueed}}, \bibinfo {author}
  {\bibfnamefont {M.~K.}\ \bibnamefont {Ma}}, \bibinfo {author} {\bibfnamefont
  {K.~A.}\ \bibnamefont {{Villegas Rosales}}}, \bibinfo {author} {\bibfnamefont
  {Y.~J.}\ \bibnamefont {Chung}}, \bibinfo {author} {\bibfnamefont {L.~N.}\
  \bibnamefont {Pfeiffer}}, \bibinfo {author} {\bibfnamefont {K.~W.}\
  \bibnamefont {West}}, \bibinfo {author} {\bibfnamefont {K.~W.}\ \bibnamefont
  {Baldwin}}, \bibinfo {author} {\bibfnamefont {J.~K.}\ \bibnamefont {Jain}}, \
  and\ \bibinfo {author} {\bibfnamefont {M.}~\bibnamefont {Shayegan}},\ }\href
  {\doibase 10.1038/s41567-020-1000-z} {\bibfield  {journal} {\bibinfo
  {journal} {Nat. Phys.}\ }\textbf {\bibinfo {volume} {17}},\ \bibinfo {pages}
  {48} (\bibinfo {year} {2021})}\BibitemShut {NoStop}%
\bibitem [{\citenamefont {Jain}(1989)}]{Jain1989}%
  \BibitemOpen
  \bibfield  {author} {\bibinfo {author} {\bibfnamefont {J.~K.}\ \bibnamefont
  {Jain}},\ }\href {\doibase 10.1103/PhysRevLett.63.199} {\bibfield  {journal}
  {\bibinfo  {journal} {Phys. Rev. Lett.}\ }\textbf {\bibinfo {volume} {63}},\
  \bibinfo {pages} {199} (\bibinfo {year} {1989})}\BibitemShut {NoStop}%
\bibitem [{\citenamefont {Park}\ and\ \citenamefont {Jain}(1998)}]{Park1998}%
  \BibitemOpen
  \bibfield  {author} {\bibinfo {author} {\bibfnamefont {K.}~\bibnamefont
  {Park}}\ and\ \bibinfo {author} {\bibfnamefont {J.~K.}\ \bibnamefont
  {Jain}},\ }\href {\doibase 10.1103/PhysRevLett.80.4237} {\bibfield  {journal}
  {\bibinfo  {journal} {Phys. Rev. Lett.}\ }\textbf {\bibinfo {volume} {80}},\
  \bibinfo {pages} {4237} (\bibinfo {year} {1998})}\BibitemShut {NoStop}%
\bibitem [{\citenamefont {Park}\ and\ \citenamefont {Jain}(1999)}]{Park1999}%
  \BibitemOpen
  \bibfield  {author} {\bibinfo {author} {\bibfnamefont {K.}~\bibnamefont
  {Park}}\ and\ \bibinfo {author} {\bibfnamefont {J.~K.}\ \bibnamefont
  {Jain}},\ }\href {\doibase 10.1103/PhysRevLett.83.5543} {\bibfield  {journal}
  {\bibinfo  {journal} {Phys. Rev. Lett.}\ }\textbf {\bibinfo {volume} {83}},\
  \bibinfo {pages} {5543} (\bibinfo {year} {1999})}\BibitemShut {NoStop}%
\bibitem [{\citenamefont {Halperin}\ \emph {et~al.}(1993)\citenamefont
  {Halperin}, \citenamefont {Lee},\ and\ \citenamefont {Read}}]{Halperin1993}%
  \BibitemOpen
  \bibfield  {author} {\bibinfo {author} {\bibfnamefont {B.~I.}\ \bibnamefont
  {Halperin}}, \bibinfo {author} {\bibfnamefont {P.~A.}\ \bibnamefont {Lee}}, \
  and\ \bibinfo {author} {\bibfnamefont {N.}~\bibnamefont {Read}},\ }\href
  {\doibase 10.1103/PhysRevB.47.7312} {\bibfield  {journal} {\bibinfo
  {journal} {Phys. Rev. B}\ }\textbf {\bibinfo {volume} {47}},\ \bibinfo
  {pages} {7312} (\bibinfo {year} {1993})}\BibitemShut {NoStop}%
\end{thebibliography}%
\end{document}